\documentstyle[11pt,newpasp,twoside,epsf]{article}
\begin{document}
\title{Discovery of X-ray emission from the proto-stellar jet L1551 
IRS5 (HH 154)}
 \author{Fabio Favata, C.\,V.\,M. Fridlund}
\affil{Astrophysics Division -- Space Science Department of ESA, ESTEC,
  Postbus 299, NL-2200 AG Noordwijk, The Netherlands}
\author{G. Micela, S. Sciortino}
\affil{Osservatorio Astronomico di Palermo, 
Piazza del Parlamento 1, I-90134 Palermo, Italy}
\author{A.\,A. Kaas}
\affil{Nordic Optical Telescope, Apartado 474, E-38700 Santa Cruz de la 
Palma, Canarias, Spain}

\begin{abstract}
  We have for the first time detected X-ray emission associated with a
  proto-stellar jet, on the jet emanating from L1551 IRS5. The IRS5
  proto-star is hidden beyond a very large absorbing column density,
  making the direct observation of the jet's emission possible. The
  observed X-ray emission is likely associated with the shock
  ``working surface'', i.e.\ the interface between the jet and the
  circumstellar medium. The X-ray luminosity emanating from the jet is
  moderate, at $L_{\rm X} \simeq 3 \times 10^{29}$ erg s$^{-1}$, a
  significant fraction of the luminosity normally associated with the
  coronal emission from young stars. The spectrum of the X-ray
  emission is compatible with thermal emission from a hot plasma, with
  $T \simeq 0.5$ MK, fully compatible with the temperature expected
  (on the basis of the jet's velocity) for the shock front produced by
  the jet hitting the circumstellar medium.
\end{abstract}

\section{Introduction}

During the final stages of the formation of low-mass stars (in the
so-called classical T~Tau phase) accretion of material from the
proto-stellar nebula onto the Young Stellar Object (YSO) takes place
through an accretion disk. Very often the presence of the accretion
disk is correlated with the presence of energetic polar outflows, that
is, collimated jets of material being ejected perpendicularly to the
disk, along its axis. When these jets collide with the surrounding
ambient medium -- or with previously ejected material -- they form a
shock structure, which is directly observable in the form of so-called
Herbig-Haro jets.

X-ray emission (and thus the presence of hot plasma, at temperatures
in excess of several $\times 10^5 K$, up to $\simeq 100$ MK during
energetic flares) has by now been observed in most stages of the
formation of low-mass stars, ranging from the highly embedded, perhaps
spherically accreting proto-stars (Class~0 objects) to the final
stages of the pre-main sequence life of a star, the Weak-Line T~Tau
stage, during which the X-ray luminosity is thought to come from a
``normal'' (however very active) stellar corona (e.g.\ Feigelson \&
Montmerle 1999).

While accretion itself has been considered as a possible source of
X-ray emission in classical T~Tau stars, up to now no evidence of
energetic phenomena associated with proto-stellar jets has been
observed. Here we present the first observations of X-ray emission
from a proto-stellar jet, obtained in a well-studied system in which
the proto-star (and its immediate circumstellar environment) powering
the outflow is so heavily obscured that the jet can be singled out as
the source of emission of the X-rays without ambiguity.  Our
observations show that this jet is indeed an X-ray source with a
luminosity equivalent to a significant fraction of the X-ray
luminosity normally associated with YSOs. The observed X-ray spectrum
is compatible with a thermal origin of the observed X-ray emission.
The associated temperature is moderate, well matched to the shock
velocities observed in this and other Herbig-Haro jets. This raises
the question of whether the X-ray emission associated with jets could
indeed be a common feature of stellar formation, so that in some cases
a significant fraction of the X-ray luminosity associated with the
star (YSO/accretion disk) is actually emanating from shocks in the
jet.

\section{The L1551 IRS5 outflow}
\label{sec:sample}

The L1551 cloud is one of the nearest ($d \simeq 140$ pc) sites of
ongoing star formation, in which objects in several different stages
of the process are clearly visible, from deeply embedded, actively
accreting (proto-)stars to the final stages of star formation
represented by the Weak-Line T~Tau stars with no remaining
circumstellar material. In this paper we are concerned with the jet
associated with the IRS5 source embedded in the L1551 cloud and its
associated outflow.

L1551 IRS5 is a deeply embedded proto-stellar binary system in the
L1551 cloud, and it is effectively invisible at optical wavelengths as
it is hidden behind some $\simeq 150$ mag of visual extinction (White
et al.\ 2000) which most likely originates in the circumstellar
accretion disk. The two Class 0/1 stars have a total luminosity of
$\approx 30\, L_\odot$, and appear to be (jointly?)  powering at least
two observable outflows, a large (several arcmin) bipolar molecular
outflow and a much smaller (with a length of $\approx 10$ arcsec)
denser two-component jet (Fridlund \& Liseau 1998), consisting of
material at temperatures of $T \simeq 10^4$ K, thus visible in the
emission lines of e.g.\ H$\alpha$.  The jet moves at transverse
velocity of 200--400 km\,s$^{-1}$ (Fridlund \& Liseau 1994) and
appears to end in a shock against the ambient medium (a ``working
surface'')

\section{XMM-Newton observations}
\label{sec:obs}

The X-ray observations presented here were obtained with the
XMM-Newton observatory. A deep (50 ks exposure) of the star-forming
region of the L1551 cloud was obtained starting on Sep.\ 9 2000 at
19:10 UTC. All three EPIC cameras were active at the time of the
observation, in full-frame mode, with the ``medium'' filters. Here we
present the data obtained with the EPIC-PN camera.

An X-ray source (the leftmost of the three sources in Fig.~1) is
clearly visible at the position of the IRS5 jet, with two additional
sources visible in the region of the molecular outflow. For each of
these sources source photons have been extracted from a circular
region of 45 arcsec diameter, while background photons have been
extracted from a region on the same CCD chip and at the same off-axis
angle as for the source region.  The spectral analysis has been
performed using the XSPEC package, after rebinning the source spectra
to a minimum of 20 source counts per (variable width) bin.  The
background-subtracted count rate for the X-ray source associated with
IRS5 is $8.4 \times 10^{-4}$ cts~s$^{-1}$ in the EPIC PN camera, so
that 42 source cts are collected the $\simeq 50$ ks exposure, allowing
only a limited amount of spectral information to be derived for the
source.  The resulting spectrum (shown in Fig.~2) is soft, and can be
reasonably described with a moderately absorbed thermal spectrum.  The
best-fit column density ($1.4 \pm 0.4 \times 10^{22}$ cm$^{-2}$)
corresponds to an extinction of $A_V = 7.3 \pm 2.1$ mag.  The best-fit
temperature is $T = 0.5 \pm 0.3 \times 10^6$~K.

The full width at half energy (FWHE) of the XMM point-spread function
(PSF) for EPIC PN camera is $\approx 15$ arcsec, significantly larger
than the size of the jets (whose visible length is $\approx 10$
arcsec).  Thus, it is not possible to locate the precise site of the
X-ray emission within the jet structure.

The two additional X-ray sources visible in Fig.~1 have spectra and
absorbing column densities compatible with their being background
active galactic nuclei shining through the molecular outflow, and are
thus most likely unrelated with IRS5 and its jet/outflow structure.

\begin{figure}
    \plottwo{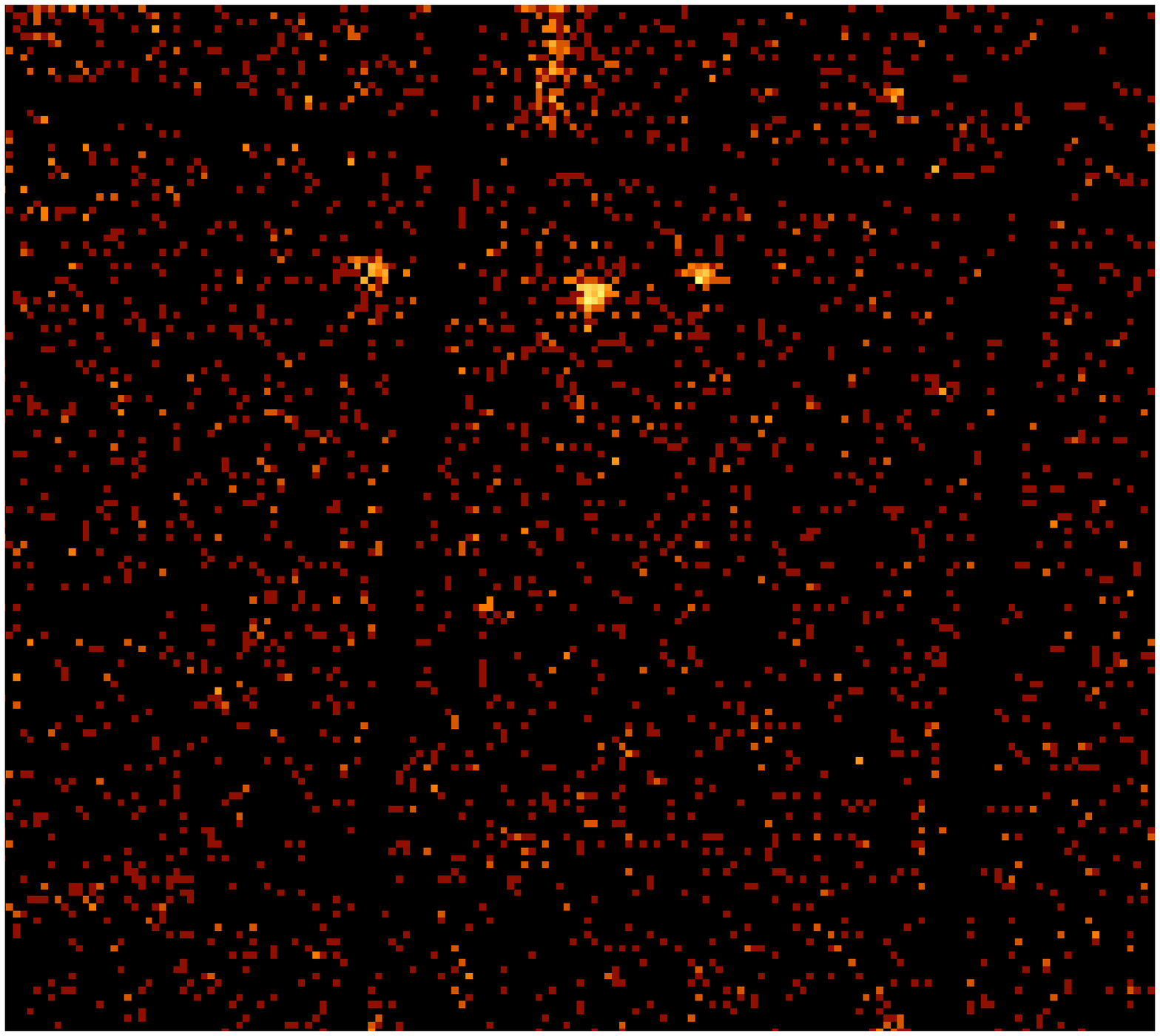}{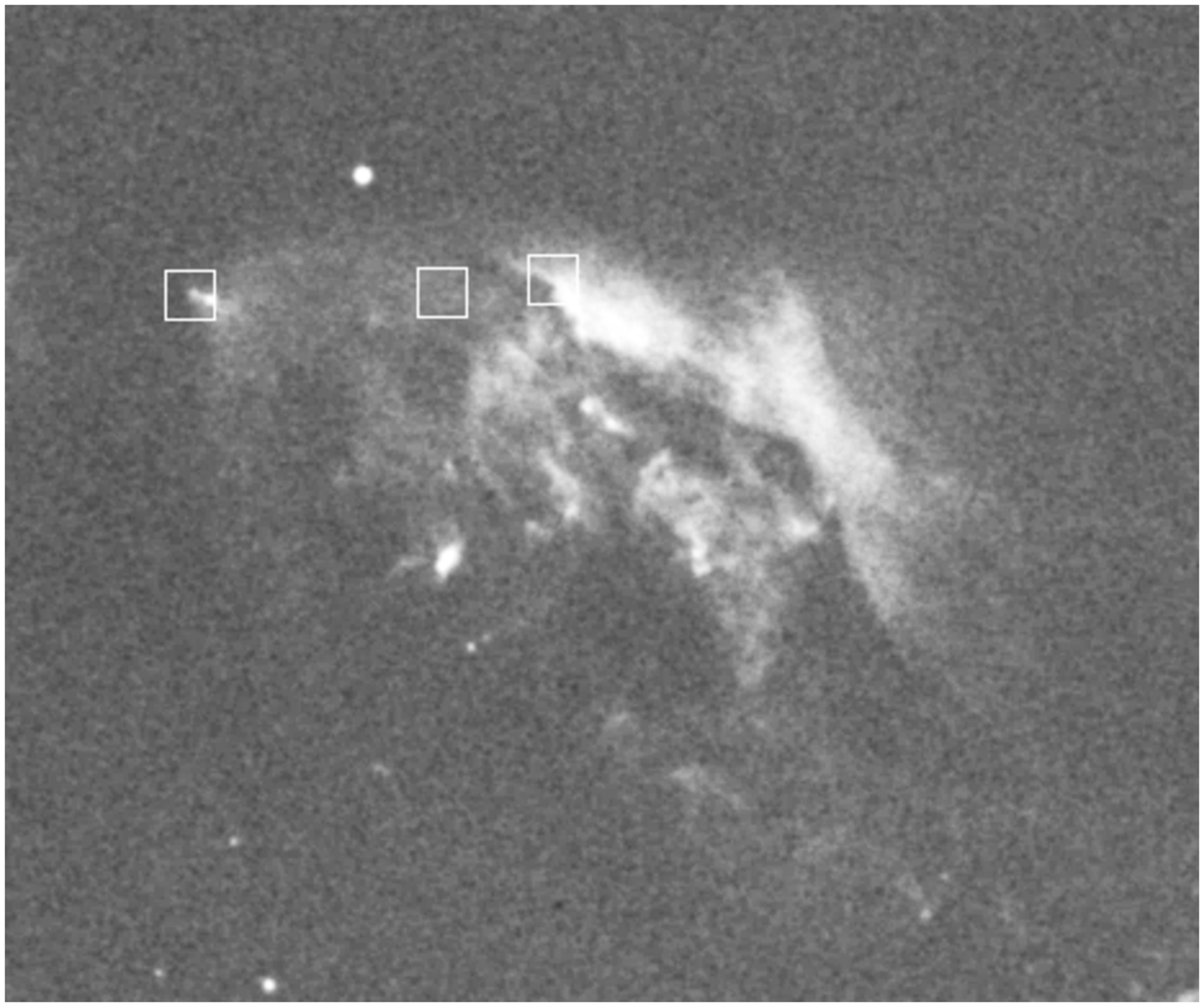}
  \caption{The left panel shows the region of L1551 in X-rays, as seen
    in the XMM EPIC-PN camera, while the right panel shows the same
    region as seen in a 300 s $R$-band CCD image obtained at the
    Nordic Optical Telescope. The two images are not exactly on the
    same scale; the position of the three X-ray sources visible in the
    left panel is indicated on the $R$-band image by the three white
    squares. The leftmost X-ray point source is the one associated
    with L1551 IRS5, while the central and rightmost point sources are
    two background sources unrelated with the jet.}
  \label{fig:image}
\end{figure}
\begin{figure}
 \plotfiddle{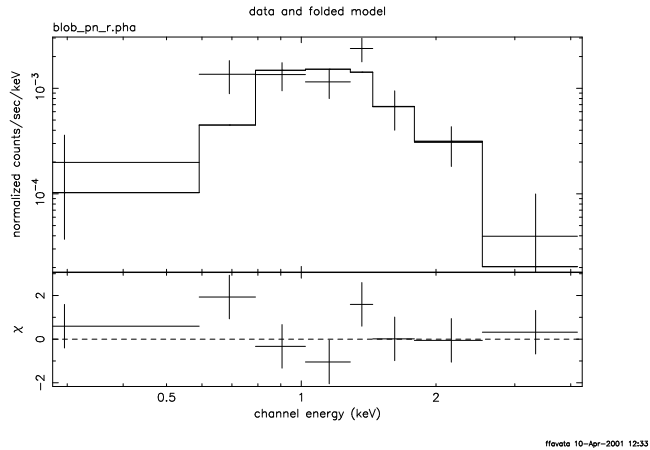}{5cm}{270}{33}{33}{-140}{190}
  \caption{The observed, background-subtracted EPIC PN X-ray spectrum
    of the X-ray source associated with the L1551 IRS5 jet. The
    best-fit thermal spectrum is also shown.}
  \label{fig:irs5spec}
\end{figure}

\section{Discussion}
\label{sec:disc}

The IRS5 jet has a number of shocks along it, and it is observed to
end in a ``working surface'' against the ambient medium at $\approx
10$ arcsec from the presumed location of the source powering it. The
absorbing column density toward the jet is estimated at 4--6 mag, a
value compatible with the X-ray measured column density, making the
association between the X-ray emission and the jet highly plausible.
Since the IRS5 proto-stellar system is hidden behind a very thick
layer of absorbing material, corresponding to $A_V \ga 150$ mag, it
can be excluded that the X-ray photons -- given the small absorbing
column density and the lack of high-energy photons in the spectrum --
emanate from (or close to) the photosphere/chromosphere of the
proto-stars powering the jet. We therefore draw the conclusion that
this source is the result of thermal emission in the shocks whose
recombination light is seen along the jet in the visual wavelength
regime.

As discussed in detail by Favata et al.\ (2001), the fluid velocity in
the shock is $\approx 270$ km\,s$^{-1}$, with an inclination,
$\mathrm{i}$, of $\approx 45$ deg. The immediate post-shock
temperature can then be estimated at $T_{\mathrm ps} \simeq 0.67$ MK,
fully compatible with the observed X-ray temperature of $0.5 \pm 0.3$
MK. Thus, the observed X-ray emission is likely to be due to material
heated at the interface shock (the working surface) between the jet
and the ambient medium medium, or possibly in shocks along the cavity
excavated by the jet.  The X-ray luminosity of the emission associated
with the jets is $L_{\rm X} \simeq 3 \times 10^{29}$ erg s$^{-1}$
(assuming a distance of 140 pc for the L1551 complex). 
% This value is of the same order as
% that of the H$\alpha$-luminosity of the brightest knot in the jet,
% which is $\simeq 4 \times 10^{28}$ erg s$^{-1}$ (Fridlund \& Liseau
% 1994).

While the jet's X-ray luminosity is at the low end of the X-ray
luminosity distribution of typical PMS stellar X-ray sources, the
location of the emitting source some 10 AU \emph{above} the star and
thus the disk may result in a relevant effect of the jet's X-ray
emission on the conditions of the disk even in the presence of a
higher stellar X-ray luminosity. The X-ray emission coming from the
jet's shock will illuminate the disk from above (and below), and can
thus change the ionization of the disk's material at large distances
from the proto-star, in regions which are normally (unless the disk is
very strongly flared) shielded from the influence of the stellar X-ray
emission.
 
\subsection{Energetics}

The mass of the jet (see Favata et al.\ 2001 for the details) can be
estimated at $1$--$2 \times 10^{-6}$~M$_{\odot}$, which, with a shock
velocity of $\approx 200$ km\,s$^{-1}$, results in a mechanical
luminosity of the jet of $10^{41}$--$10^{42}$ erg s$^{-1}$, so that a
very low conversion efficiency between mechanical and radiant
luminosity is sufficient to justify the observed X-ray luminosity from
the shock.  The H$\alpha$ luminosity is $\approx 4.7 \times 10^{28}$
erg s$^{-1}$, comparable with the X-ray luminosity derived here.

\section{Conclusions}

While the energetic nature of the collimated jets observed to be
originating from proto-stellar sources has been evident for some time,
no high-energy photons have up to now been observed from these
phenomena. Here we report the first convincing evidence of X-ray
emission from the proto-stellar jet associated with the IRS5
proto-star(s) in the L1551 cloud. The X-ray source and the proto-star
and related jets are positionally coincident, and the small absorbing
column density observed for the X-ray spectrum ($\simeq 7$ mag, fully
compatible with the absorbing column density observed towards the jet)
allow us to exclude that the X-ray emission is associated with the
proto-stellar sources (which are hidden behind $\approx 150$ mag of
obscuration). The size of the jets ($\approx 10$ arcsec) originating
at L1551 IRS5 is smaller in angular extent than the XMM EPIC PSF
($\approx 15$ arcsec), so that no inference is possible on spatial
grounds about the possible detailed location of the origin of the
X-ray emission.

The emission from the IRS5 jet is compatible with being caused by
thermal emission from a plasma heated to a moderate temperature ($T
\simeq 0.5$~MK). This is equivalent to the shock temperature that is
expected at the interface (``working surface'') between the jet and
the surrounding circumstellar medium, on the basis of the observed jet
velocity.

\end{document}